\documentclass[12pt]{iopart}
\usepackage{graphicx}
\usepackage{times}
\begin{document}

\title{Are moving punctures equivalent to moving black holes?}

\author{Jonathan~Thornburg$^{1,2}$, Peter~Diener,$^{3,4}$,
  Denis~Pollney$^{1,3}$, Luciano~Rezzolla$^{1,4}$,
  Erik~Schnetter$^{3,4}$, Ed Seidel$^{3,4}$, Ryoji~Takahashi$^{3,5}$}

\address{$^{1}$
  Max-Planck-Institut f\"ur Gravitationsphysik,
  Albert-Einstein-Institut,
Potsdam, Germany
}
\address{$^{2}$
  School of Mathematics,
  University of Southampton,
  Southampton, England
}

\address{$^{3}$
  Center for Computation \& Technology,
  Louisiana State University,
  Baton~Rouge, LA, USA
}
\address{$^{4}$
  Department of Physics and Astronomy,
  Louisiana State University,
  Baton~Rouge, LA, USA
}

\address{$^{5}$
  Instituto de Ciencias Nucleares, Universidad Nacional Aut\'onoma de
  M\'exico, M\'exico D.F., M\'exico}


\begin{abstract}
	When simulating the inspiral and coalescence of a binary
	black-hole system, special care needs to be taken in handling
	the singularities.  Two main techniques are used in
	numerical-relativity simulations: A first and more traditional
	one ``excises'' a spatial neighbourhood of the singularity from
	the numerical grid on each spacelike hypersurface. A second
	and more recent one, instead, begins with a ``puncture''
	solution and then evolves the full 3-metric, including the
	singular point.  In the continuum limit, excision is justified
	by the light-cone structure of the Einstein equations and, in
	practice, can give accurate numerical solutions when suitable
	discretizations are used. However, because the field variables
	are non-differentiable at the puncture, there is no proof that
	the moving-punctures technique is correct, particularly in the
	discrete case. To investigate this question we use both
	techniques to evolve a binary system of equal-mass
	non-spinning black holes.  We compare the evolution of two
	curvature 4-scalars with proper time along the
	invariantly-defined worldline midway between the two black
	holes, using Richardson extrapolation to reduce the influence
	of finite-difference truncation errors. We find that the
	excision and moving-punctures evolutions produce the same
	invariants along that worldline, and thus the same spacetimes
	throughout that worldline's causal past. This provides
	convincing evidence that moving-punctures are indeed
	equivalent to moving black holes.
\end{abstract}

\pacs{
04.25.Dm, 
04.30.Db, 
04.70.Bw, 
95.30.Sf, 
97.60.Lf  
}

\section{Introduction and Motivations} 

Binary black hole coalescences are both natural laboratories in which
to study the nonlinear strong-field dynamics of General Relativity and
among the most promising sources of gravitational radiation for modern
laser-interferometric detectors. Despite these being very simple
systems, as the black holes are assumed to be in vacuum and the
solution of the Einstein equations fully describes the binary, no
analytic solutions are known and numerical methods represent the only
viable approach to investigate these systems' strong-field dynamics.
The past
few years have seen major advances in these numerical simulations,
with demonstrations of multiple orbit evolutions through
merger~\cite{Pretorius:2005gq, Pretorius:2006tp, Baker:2006yw,
Campanelli:2006gf, Baker:2006ls}, recoils from unequal-mass
systems~\cite{Baker:2006nr, Gonzales06tr}, and studies of spin
couplings in the final orbit~\cite{Campanelli:2006uy,
Campanelli:2006vp, Campanelli:2006fy}. Convergence studies and
cross-checks between independent codes~\cite{Baker-etal:2007a} have
demonstrated an impressive consistency, lending support to their
credibility as reliable modellers of these sources.

Any such numerical simulation must use some means to treat the
singularities contained within the black holes and modern simulations
therefore use different techniques to treat the black holes, either
``excision'' or ``moving punctures''. The excision
technique~\cite{Seidel92a} can use a slicing which intersects the
singularity, but removes part of the interior of the horizon from the
numerical domain on each slice. Excision is straightforward in
spherical symmetry, but technically more difficult to implement in
higher dimensions for grids using Cartesian coordinates, in which the
excision region is an irregular surface of spherical topology.  The
simplest case is that of ``stationary'' excision, where once a given
grid point is excised, it stays excised for the remainder of the
numerical simulation. For an orbiting binary black hole system this
requires using coordinates which corotate with the black holes. This
technique has been used by several authors (see,
\textit{e.g.},~\cite{Alcubierre00a, Bruegmann:2003aw,
Alcubierre2003:pre-ISCO-coalescence-times, Diener-etal-2006a}) but
we have so far not been able to compute useful waveforms from inspiral
simulations using this technique. In contrast, a technically more
difficult form of excision allows the excision region to move with
respect to the numerical grid. This technique has also been used by
several authors (see, \textit{e.g.},~\cite{Shoemaker2003a,
Sperhake:2003fc, Sperhake2005a, Pretorius:2005gq, Pretorius:2006tp,
Szilagyi-etal-2006a}) and in some cases has allowed the calculation of
waveforms~\cite{Pretorius:2005gq, Pretorius:2006tp,
Szilagyi-etal-2006a}.

The ``moving puncture'' technique, on the other hand, makes use of
``puncture data''~\cite{Brandt97b} which are evolved \emph{without
excision} using suitable gauges~\cite{Alcubierre02a} and allowing the
singularities to be advected across the computational
grid~\cite{Baker:2005vv, Campanelli:2005dd}. We recall that by this
method the curvature singularity at the centre of a black hole is
avoided and replaced by an asymptotically flat spacetime through the
throat. A coordinate singularity at the effective $r=0$ of each black
hole still remains, and this represents a non-differentiable point
which, at least in principle, needs special treatment. Standard
finite-difference techniques, in fact, require smooth functions at
each gridpoint and thus would not be able to evaluate derivatives in
the neighbourhood of the puncture. In practice, however, the
inaccuracies at these points are isolated and, at least in the
continuum limit, the physical causality of the spacetime ensures that
these errors do not propagate out of the horizon. In addition, the
standard singularity-avoiding gauge conditions used in puncture
evolutions lead to spacetimes that are essentially stationary in their
neighbourhood. This has been pointed out in~\cite{Baiotti06} and more
extensively discussed in~\cite{Hannam:2006vv}.

An important remark should be made at this point. Mathematically,
either use of the excision or moving-punctures technique is justified
by the light-cone causality of the Einstein equations near a black
hole, which guarantee that within the horizon physical modes only
propagate inwards, towards the spacelike excision boundary or the
puncture. In practice, there are two factors which complicate this
picture. Firstly, the conformal-traceless formulations of the Einstein
equations~\cite{Nakamura87, Baumgarte99, Alcubierre99d} evolution
system, with the commonly used gauges, are known to have gauge modes
which propagate superluminally in the neighbourhood of the black
hole~\cite{Alcubierre99e}. And secondly, it is only in the
\textit{continuum form} that the characteristic structure completely
determines the causality. For the \textit{discretized form} of the
Einstein equations, numerical errors having high spatial frequencies
with respect to the grid spacing are inevitably generated. Such
signals are not propagated accurately by a finite-differencing scheme
and, in particular, they are not constrained to propagate within the
light cone; indeed they can propagate at any speeds up to the
finite-difference domain of dependence speed. It is thus not obvious
that such spurious modes will remain confined within the black hole
horizon. This concept is so essential for understanding this work that
we will stress it again: there are as yet no rigorous proofs that
either excision or moving-punctures techniques yield stable evolution
schemes for the conformal-traceless formulations of the Einstein
equations, or that if so, that the results will converge to a
(correct) solution of the Einstein equations as the grid is refined.

The purpose of this paper is to improve our confidence in both methods
in situations that go beyond simple static or stationary solution of
isolated black holes. We do this by comparing the spacetimes generated
using corotating-excision (CE) and moving-punctures (MP) evolutions of
the same initial data, representing an equal-mass non-spinning binary
black hole in its last orbit before coalescence. In particular, we
concentrate on the evolution of two curvature invariants measured
along a well-defined geodesic between the two black holes and provide
the first strong-field evidence that excised and moving-puncture yield
the same solution of the Einstein equations for this system.


\section{Methods and Results} 

The simplest way to compare evolutions of binary systems using either
CE or MP techniques would involve the direct use of gauge-invariant
quantities such as waveforms. Indeed, this has been done
in~\cite{Alcubierre2003:BBH0-excision}, for head-on collisions of pure
puncture evolutions, as well as in~\cite{Baker-etal:2007a}, where
waveforms coming from different implementations of the moving puncture
technique and from a generalized harmonic formulation of the Einstein
equations~\cite{Pretorius:2006tp} were compared. While both of these
works have shown there are close similarities in the waveforms, they
have also highlighted small differences. Most importantly, however,
the comparisons were not using Richardson extrapolation to reduce
the influence of finite-difference truncation errors. Unfortunately,
because of technical complications in the wave-extraction when using
corotating coordinates, the evolutions of corotating and excised
punctures have not produced usable asymptotic
waveforms~\cite{Bruegmann:2003aw, Diener-etal-2006a}
(\cite{Scheel-etal-2006:dual-frame} presents a possible route to
overcoming these problems). Thus, the use of waveforms is not a
viable route for this comparison.

However, an alternative route, which also allows to probe regions of
the spacetime with strong and highly dynamical curvature, consists of
first identifying corresponding events in the two spacetimes, and then
gauge-transform quantities that are gauge-dependent. (Note that
similar issues arise in almost any comparison of different
numerical-relativity codes~\cite{Choptuik-Goldwirth-Piran-1992}.). For
an equal-mass binary system this is particularly simple and to
identify corresponding events we consider the central worldline midway
between the two black holes. The initial data has $\pi$-symmetry about
this worldline, and this is preserved by both evolutions, so this
worldline is invariantly defined. We can thus use proper time along
this worldline as a 4-invariant parameterization, matching up
corresponding events in the two simulations.

All the numerical simulations for both corotating excision and moving
punctures have been performed using the same evolution code and
initial data. The latter, in particular, are constructed as
in~\cite{Ansorg:2004ds} and have orbital parameters to approximate a
binary system of non-spinning black holes in quasi-circular orbit,
with initial proper separation $L \,{=}\, 9.32M$, mass parameters $m
\,{=}\,0.47656M$, where $M$ is the total mass of the system, and equal
and opposite linear momenta $p \,{=}\,\pm
0.13808M$~\cite{Bruegmann:2003aw}. The evolutions are carried out
using a conformal-traceless formulation of the Einstein equations as
described in~\cite{Alcubierre02a}, with ``$1{+}\log$'' slicing and
$\Gamma$-driver shift. The CE runs benefit from insights gained
in~\cite{Diener-etal-2006a} and use the GC2~gauge condition of that
work.  The MP runs use the optimal gauge conditions
of~\cite{vanMeter:2006vi}, with the lapse evolved via
$\partial_t{\alpha} = -2\alpha K+\beta^{i}\partial_{i}\alpha$, while
the shift evolution follows prescription~8 in Table~I
of~\cite{vanMeter:2006vi} with $\eta = 0.5$. Individual apparent
horizons are located every few timesteps during the
evolution~\cite{Thornburg95,Thornburg2003:AH-finding}.
The code is implemented in the Cactus framework.

Spatial differentiation is performed via straightforward
finite-differencing using second- or fourth-order algorithms for CE
and MP, respectively. In addition, for the MP runs a fifth-order
Kreiss-Oliger artificial dissipation is also added to all evolution
variables.  Vertex-centred AMR is employed using nested mesh-refined
grids~\cite{Schnetter-etal-03b} with the highest resolution
concentrated in the neighbourhood of the individual horizons.  In the
case of CE evolutions, eight levels of refinement have been used; the
corotating gauge conditions guarantee that the black holes remain on
the fine grids throughout the evolution.  In the case of MP
evolutions, on the other hand, nine levels of refinement are used,
with the finest two levels being locked to the position of the
centroid of the apparent horizon. For either the CE or MP approach, we
have carried out simulations with at least three different
resolutions.  However, because the two approaches have rather
different truncation errors, with MP using higher-order finite differencing,
the CE simulations have fine-grid spatial resolutions of
$h=0.018$, $0.015$, and $0.0125\,M$, while the MP ones have coarser
resolutions, with $h=0.032$, $0.025$, and $0.020\,M$.

As mentioned earlier, an unambiguous measure of the CE and MP spacetimes
can be made by using the 4-invariant spacetime curvature scalars $I
\,{\equiv}\, \tilde{C}_{\alpha\beta\gamma\delta}
\tilde{C}^{\alpha\beta\gamma\delta}$ and $J \,{\equiv}\,
\tilde{C}_{\alpha\beta\gamma\delta} \tilde{C}^{\gamma\delta}{}_{\mu\nu}
\tilde{C}^{\mu\nu\alpha\beta}$, where
$\tilde{C}_{\alpha\beta\gamma\delta} \,{=}\, C_{\alpha\beta\gamma\delta}
+ \frac{1}{2} i \epsilon_{\alpha\beta\mu\nu} C^{\mu\nu}{}_{\gamma\delta}$
is the self-dual part of the Weyl tensor $C_{\alpha\beta\gamma\delta}$.
Note that while $I$ and $J$ are complex numbers, for our evolutions their
real parts are at least 12 orders of magnitude larger than the imaginary
ones, so that $I,\,J=\Re(I,\,J)$ to very good precision.  Hereafter we
will concentrate on reporting results for $I$ only, as a very similar
behaviour is found also for $J$.

\begin{figure}[tbp!]
\centerline{
\resizebox{8.5cm}{!}{\includegraphics[angle=-0]{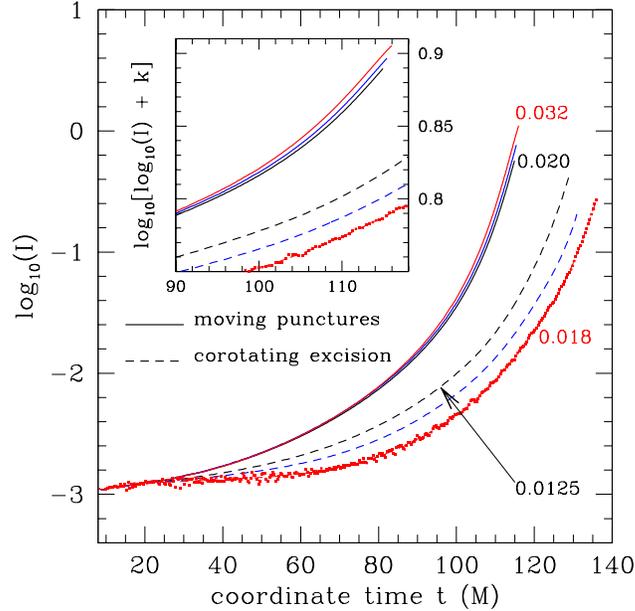}
}}
\caption{Evolution in coordinate time of $\log I$ for the MP
  evolutions (solid lines) and the smoothed CE ones (dashed lines),
  while the raw CE-data is indicated with squares for the coarsest
  resolution only. A magnification of the overlapping MP curves is
  shown in the inset.}
\label{fig:I_vs_t}
\end{figure}

The measure of the invariants has to be made along exactly the same
worldine in the two spacetimes, and when the black holes have equal
masses, the spatial point midway between the black holes is
invariantly defined, and its worldline can be used for this
measure. Clearly, evolutions with different gauges will generate
different coordinate descriptions of this point, but this ambiguity is
absent when the affine parameter along the geodesic is chosen to be
the proper time $\tau$. As a result, $I$ expressed as a function of
$\tau$ along the worldline of the midway point between the two black
holes can be used as a gauge-invariant diagnostic of the evolution. We
note that because $I(\tau)$ has a super-exponential behaviour, all of
the analysis has been performed in terms of $\log I$ to increase
accuracy.

Figure~\ref{fig:I_vs_t} shows the evolution in coordinate time of $\log
I$ for the MP evolutions (solid lines) and the smoothed CE ones (dashed
lines). The raw $I$-timeseries for the CE evolutions are quite noisy
(cf.\ small squares in Fig.~\ref{fig:I_vs_t}), and for further analysis
we smoothed these with a fourth-order Savitzky-Golay
filter~\cite{Savitzky-Golay-1964} over a $\pm 10\,M$ sliding window in
coordinate time.  We have verified the smoothing does not introduce
systematic errors; no smoothing was necessary for the MP evolutions.

Because it is not practical to make $h$ small enough so that
finite-differencing errors are negligible, we exploit the known
convergence properties of finite-difference schemes to
Richardson-extrapolate our finite-$h$ results to the limit $h \to
0$. In particular, given some quantity~$u$ computed at numerical
resolution~$h$, we write the Richardson-extrapolation series $u(h)$ as
$u(h) = u(0) + p h^n + q h^{n{+}1} + {\cal O}(h^{n+2}) \;,$
%
where $n \,{=}\, 2\, (4)$ for CE~(MP), and where the coefficients $p$
and $q$ depend on $u$, but not on the resolution~$h$.  Given $u(h)$ at
three distinct resolutions, we solve for $u(0)$ as the
Richardson-extrapolated value for $u$, i.e., ${\cal R}(u) \equiv
u(0)$. Clearly, slightly different values for ${\cal R}(u)$ will be
obtained depending on which of the higher-order terms are neglected in
the series expansion,
%
%
and we use the magnitude of the last known term in the expansion
%
%
at the highest resolution as a rough estimate of the errors in ${\cal
R}(u)$.

\begin{figure}[tbp!]
\centerline{
\resizebox{8.5cm}{!}{\includegraphics[angle=-0]{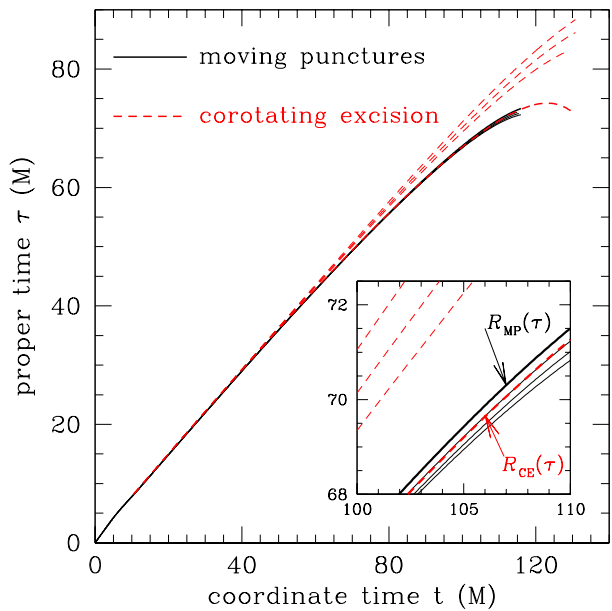}
}}
\caption{Relationship between coordinate time~$t$ and proper
  time~$\tau$ for simulations using MP (solid lines) or CE (dashed
  lines). The inset offers a magnification over a representative
  window in time.}
\label{fig:tau_vs_t}
\end{figure}

In practice, for each evolution we have first extracted the timeseries
of $\alpha$ and $I$ up to the detection of a common apparent horizon
and then time-integrated $\alpha(t)$ to obtain $\tau(t)$, as shown in
Fig.~\ref{fig:tau_vs_t} for simulations using MP (thin solid lines) or
CE (thin dashed lines).  Using this data and the Richardson-extrapolation
series expansion
%
%
, an estimate for ${\cal R}(\tau(t))$ is then
obtained and shown with thick lines (solid for MP and dashed for CE),
with the inset offering a view.  Despite having lower resolutions, the
MP evolutions show a closer match between the different resolutions
and the Richardson-extrapolated result than do the CE ones.

Finally, we have Richardson-extrapolated $\log I(t)$, and removed the
dependence on the time coordinate by mapping $t$ to ${\cal R}\bigr(
\tau(t) \bigr)$ (cf.~Fig.~\ref{fig:tau_vs_t}).  Our end results are
therefore ${\cal R} \bigl( \log I_{_{\rm CE}} \bigr)$ and ${\cal R}
\bigl( \log I_{_{\rm MP}} \bigr)$, both as functions of ${\cal
R}(\tau)$.

\begin{figure}[tbp!]
\centerline{
\resizebox{8.5cm}{!}{\includegraphics[angle=-0]{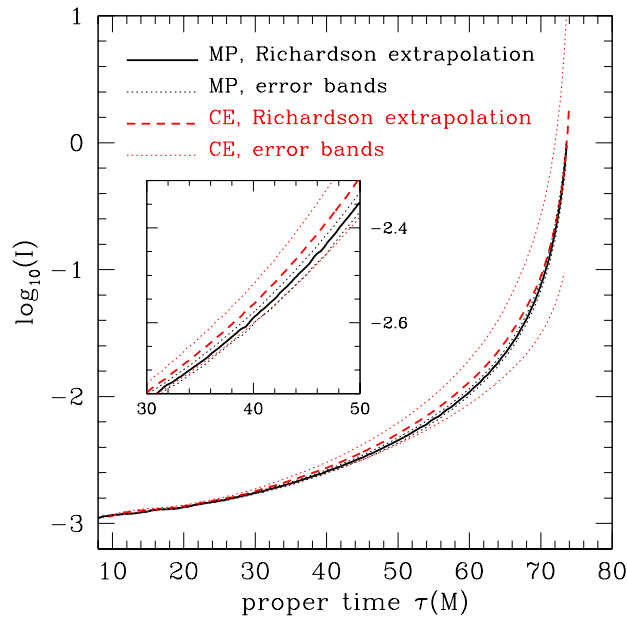}
}}
\caption{$\log I(\tau)$ for each evolution family, together with the
  estimated errors. Thick lines show the Richardson-extrapolated
  results (solid for MP and dashed for CE) while the dotted lines
  report the error bars, with the larger ones referring to CE
  evolutions. Note the excellent agreement as highlighted in the
  inset.}
\label{fig:I_vs_tau}
\end{figure}

The results of this procedure are summarized in
Fig.~\ref{fig:I_vs_tau}, which shows the proper-time evolution of
$\log I(\tau)$, together with the estimated error bands. More
specifically, thick lines show the Richardson-extrapolated results
(solid for MP and dashed for CE) while the dotted lines report the
error bars, with the larger ones referring to CE evolutions.  Clearly,
the two Richardson-extrapolated evolutions of the invariant lie well
within the estimated error-bands for both evolutions and are almost
indistinguishable for large portions of the simulations, despite the
large dynamical range. The inset highlights this, with a view in a
representative window in proper time.  Overall, the results in
Fig.~\ref{fig:I_vs_tau}, together with the similar ones for $J$,
demonstrate that, despite the different gauges and the different way
in which the singularities are treated in the two approaches, the two
approaches are indeed converging to the same spacetime, at least along
the fiducial central geodesic.

\section{Conclusions} 

Excision and moving punctures both seem to work well in practice, and
are used by multiple research groups, yet lack rigorous mathematical
correctness proofs.  To help improve our confidence in both methods,
we have compared them by evolving the same equal-mass non-spinning
binary black hole system using both corotating-excision and
moving-punctures techniques.  Comparing the evolution of the $I$ and
$J$ curvature 4-scalars with proper time along the invariantly-defined
worldline midway between the two black holes, and using Richardson
extrapolation to reduce the effects of finite-difference truncation
errors, we find that moving-punctures and excision evolutions agree to
within our estimated numerical errors.

$I$ and $J$ are sensitive and nonlinear functions of many components
of the Riemann tensor with markedly different nonlinearities: $I$ is
quadratic in the Riemann tensor, while $J$ is cubic. The fact they
both agree in the two different techniques makes it very unlikely that
this is just an artifact of the symmetry. A rigorous proof of the
equivalence of the two spacetimes would require a detailed examination
of the curvature components and their derivatives in an invariant
frame~\cite{Karlhede:1979ri}. However, the established equivalence of
$I$ and $J$ is a strong validation that the entire Riemann tensors for
this algebraically general spacetime agree along the central
worldline. Given the causal structure of the Einstein equations, this
agreement extends to the entire causal past of the central worldline.
Furthermore, because our data span a time much longer than the initial
separation of the two black holes (our evolutions last for $\sim
70\,M$, while the initial separation is $\sim 9\,M$), the causal past
of central worldline includes a large part of the strong-field region
of the spacetime extending well out into the wave zone. Therefore, for
the evolutions reported here these results provide convincing evidence
that the corotating-excision and the moving-punctures techniques yield
the same spacetime as solution of the Einstein equations.


\ack

The numerical calculations were performed on \textit{Peyote} and
\textit{Belladonna} at AEI, \textit{Jacquard} at NERSC,
\textit{Tungsten} at NCSA, \textit{Supermike} and \textit{Santaka} at
LSU and on \textit{Ducky} and \textit{Neptune} at LONI. This work was
supported in part by the DFG grant SFB TR/7 and by the CCT at LSU.


\section*{References}

\bibliographystyle{iopart-num-noeprint}



\bibliography{references}

\end{document}